\documentclass[usenatbib]{mnras}

\usepackage{txfonts}

\usepackage[T1]{fontenc}
\usepackage{ae,aecompl}

\usepackage{graphicx}	
\usepackage{amssymb}	
\usepackage{relsize}    

\def\apj{ApJ}

\def\apjs{ApJ Suppl. Ser.}

\def\aap{A\&A}
\def\aaps{A\&A Suppl. Ser.}

\def\araa{ARAA}

\def\mnras{MNRAS}
\def\nat{Nature}

\def\zap{ZA}

\def\beq#1{\begin{equation}\label{#1}}
\def\eeq{\end{equation}}
\def\beqa#1{\begin{eqnarray}\label{#1}}
\def\eeqa{\end{eqnarray}}

\def\apgt{\ {\raise-.5ex\hbox{$\buildrel>\over\sim$}}\ }
\def\aplt{\ {\raise-.5ex\hbox{$\buildrel<\over\sim$}}\ }
\def\lb1{LB-1}      

\newcommand{\mbh}{${\mathrm M_{BH}}$}

\newcommand{\mvis}{${\mathrm M_{vis}}$}

\newcommand{\ace}{\mbox {$\alpha_\mathrm{ce}$}}
\newcommand{\ms}{$M_\odot$}
\newcommand{\msun}{$M_\odot$}

\newcommand{\kms}{\:\mbox{${\mathrm{ km\, s^{-1}}}$}}
\newcommand{\porb}{\mbox {$P_{\rm orb}$}}

\newcommand{\teff}{\mbox{$T_{\mathrm eff}$}}

\title[Stripped He-stars: 
 the case of LB-1]{Galactic population of black holes 
 in detached binaries with low-mass stripped helium stars:
 the case of LB-1~(LS~V+22~25)}

\author[L.R. Yungelson et al.]{L.R. Yungelson,$^{1}$\thanks{E-mail: lev.yungelson@gmail.com}
A.G. Kuranov,$^2$
K.A. Postnov,$^{2,3}$ 
D.A. Kolesnikov$^2$\\
$^{1}$ Institute of Astronomy, Russian Academy of Sciences, 48 Pyatnitskaya str., 119017 Moscow, Russia\\
$^{2}$ Sternberg Astronomical Institute, M.V. Lomonosov Moscow State University, 13 Universitetskij pr., 119234 Moscow, Russia\\
$^{3}$ Kazan Federal University, Kremlevskaya 18, 420008 Kazan, Russia
}
\date{Received April 10, 2020; accepted April 23, 2020; in original form March 18, 2020}

\pubyear{2020}

\begin{document} 
\label{firstpage}
\pagerange{\pageref{firstpage}--\pageref{lastpage}}

\maketitle

\begin{abstract}
We model the Galactic population of  detached binaries that harbour black holes with 
(0.5-1.7)\,\msun\ companions -- remnants of case B mass 
exchange that rapidly cross Hertzsprung gap after the termination of the Roche-lobe overflow 
or as He-shell burning stars. Several such binaries can be currently present in the Galaxy.
The range of \mbh\ in them is about 4 to 10 \ms, the orbital periods are tens to hundreds day.
The unique BH-binary LB-1 fits well into this extremely
rare class of double stars.   
\end{abstract}



\section{Introduction}
\label{sec:intro}
\citet{2019Natur.575..618L} reported the discovery of a unique Galactic
long-period (\porb=$78.9\pm0.3$~day) spectroscopic binary \lb1\ (LS~V+22~25)
harbouring a B-type star with an invisible massive compact companion. They
associated broad H$\alpha$ emission line observed in the spectrum with the
invisible component and, based on this, estimated the masses of components as
$8.2^{+0.4}_{-1.2}$\,\msun\ and $68^{+11}_{-13}$\,\msun, respectively. A
$\simeq$70\,\ms\ compact object should be a black hole (BH). However, currently
it is expected that there exists a gap in the masses of Galactic BHs  between about
40 and 120\,\msun\ due to the pair-instability phenomenon \citep[see, e.g.,
][and references therein]{2019ApJ...887...53F,2020MNRAS.493.4333R}.

\citet{2019arXiv191204092A} and \citet{2020MNRAS.493L..22E} questioned the
association of the periodic variability of the H{$\alpha$} emission with the
compact object and suggested that it can be related to the visible star or
circumbinary matter. Also, atmospheric parameters of the B-component inferred
from spectroscopic measurements were reanalysed using new and archival data
\citep[e.g.,][]{2020A&A...633L...5I,2020MNRAS.493L..22E,2020A&A...634L...7S}.
These revisions resulted in the reduction of the implied mass of the compact
object to about (4 - 20)\,\ms. Besides, \citet{2019arXiv191112581S} have shown
that implied \mbh$\approx$70\ms\ can not result from the merger of two less
massive BH components in a triple system, while according to
\citet{2019arXiv191204509T} \lb1\ can not be formed by dynamic processes in a
cluster.

\citet{2020A&A...633L...5I}, based on a quantitative spectroscopic analysis,
photometric measurements of the angular diameter of the system and its
\textit{Gaia} parallax, argued that the visible star mass is $1.1\pm0.5$\,\msun\
and it may be a ``stripped helium star of sdB-type, masquerading as a normal
B-star; the mass of the compact object in \lb1\ exceeds
$2.5^{+0.4}_{-0.5}$\,\msun''. If the orbital inclination is not too extreme ($i
\gtrsim 20^\circ$), the mass of the compact object can be $\aplt 20$\,\msun.
Importantly, spectroscopic analysis
\citep{2020A&A...633L...5I,2020A&A...634L...7S} revealed that the matter in the
surface layers of the visible component is
CN-processed\footnote{\citet{2019Natur.575..618L} claim that the visible
companion is best-fit by a B-subgiant located on the HRD close to the
main-sequence turn-off. However, its spectrum is definitely neither that of an
sdB star nor of a normal main-sequence B-type star because of the signal of the
CNO-processed material (U. Heber, private communication).}.

The mass $1.1\pm0.5$\,\msun\ exceeds ascribed to subdwarfs 
``canonical'' mass 0.47\,\msun\ \citep[see ][for a review]{2009ARA&A..47..211H}.  ``Stripped helium stars''  are  considered as the 
remnants of donors in ``case B'' Roche lobe overflow (RLOF), which occurs
after hydrogen exhaustion, but prior to helium ignition in the stellar core
\citep{1967ZA.....65..251K}. Hence, the remnant masses should range from the minimum mass for core helium ignition, $\simeq$ 0.33\,\msun, to several dozens \msun. 
However, to the best of our knowledge, there are only five helium subdwarfs with measured or estimated masses 
between 0.6 and 1.7\,\msun, comparable to the estimate for visual component of LB-1.  
Between the most massive known subdwarf, 60~Cyg,  \citep[1.7\,\msun,][]{2017ApJ...843...60W} and the least massive Galactic WR stars
\citep[$\sim$7\,\msun, e.g., ][]{2020A&A...634A..79S}
no He-stars have been observed for as yet unknown reason.

\citet{2020A&A...633L...5I} found that the optical component of \lb1\ has
$T_{eff} = 12\,720\pm 260$~K and $\log(g) = 3.00\pm0.08$, which is consistent
with \citet{2019arXiv191204092A} data, but differ from
\citet{2019Natur.575..618L}: $T_{eff} = 18\,104 \pm 825$\,K, $\log(g) = 3.43 \pm
0.15$. However, the estimated $T_{eff}$ and $\log(g)$ are significantly lower
than the typical ones for sdB stars. 

\citet{2019Natur.575..618L} and \citet{2020A&A...633L...5I} noted that the
visual star in \lb1\ may be in the state after completion of the
RLOF episode in the red giant (RG) region on the Hertzsprung-Russel
diagram (HRD), when it crosses HRD in the Kelvin-Helmholtz time
scale toward high $T_{eff}$. Independently, \citet{2019arXiv191203599E}, ran a
grid of evolutionary tracks by a binary population synthesis (BPS) code 
and found several systems with stripped helium components that fit within
certain accuracy the parameters of \lb1; the best matching systems they found
have \mbh$\approx$7\,\ms\ and \mvis $\approx$1\,\ms; the optical components are
in the H-shell burning stage.  

We aim at verifying whether the standard stellar binary evolution is able to
reproduce the mass, effective temperature, luminosity of the optical star in
\lb1, and the ``typical'' stellar mass of BH-component. We model the population
of similar binaries and evaluate the Galactic number of \lb1\ type systems. To
this end, we use the population synthesis  and supplementary evolutionary
calculations.

\section{The model}
\label{sec:model}

Motivated by the relatively low values of \teff\ and $\log$(g) of the optical
component in \lb1, we try to find detached binaries in which (i) the initially
more massive component has evolved into a BH, (ii) the binary system has
survived the SN explosion, and (iii) the companion to the BH has completed case
B mass exchange and either crosses HRD toward high \teff\ having a He-core and a
H-burning shell or has a CO-core and a He-burning shell and evolves toward lower
\teff\ (or makes loops in HRD). We select the binaries with \mvis=(0.5 -
1.7)\,$M_\odot$. We do not consider stripped He-core burning stars as viable
components of \lb1\ stars, because of their high $\log(g)$. 
We model the population of binaries with components similar to
those of \lb1, irrespective of their orbital periods and luminosity and provide
an example of a binary which reasonably fits all parameters of \lb1.

As the first step, we apply an updated version of the open BPS code BSE
\citep{htp02} to model a population of binaries in which the primary component
has collapsed into BH and the secondary companion is a moderate-mass (4 - 9
\msun) star which has evolved into a He-star after case B RLOF. Such binaries
are expected to have passed through an evolutionary stage where they could
resemble \lb1, or to become similar to \lb1\ later\footnote{The lower boundary
of the compact star mass in \lb1\ \citep{2020A&A...633L...5I} does not exclude a
massive neutron star. However, the Galactic number of NS+stripped He star
binaries was found to be insignificant.}. 

Our typical BPS run traces the evolution of $1.5\times10^5$ binaries.
The initial masses of the primary components in the binaries are distributed according to the Salpeter's law. 
The binary mass ratios $q=M_2/M_1\le 1$ and orbital eccentricities 
$e$ are distributed as $dN/dq$=const., $q\in$\,[0.1:1] and 
$dN/de$=const. $e\in$\,[0,1]. 
Galactic stellar binarity rate $B$ is set to 50\%,
i.e., 2/3 of all stars 
reside in binaries with assumed metallicity Z=0.02.
The initial distribution of the binary orbital periods is taken as    
$f(\log \porb) \propto \log \porb^{-0.55}$ \citep{2012Sci...337..444S}. 
Stellar winds of massive stars and stripped He-stars are treated as in  \citet{2001A&A...369..574V} and \citet{2017A&A...607L...8V}.

BHs are assumed to form by direct collapse of the progenitor's CO-core, i.e., the mass of a nascent BH is set equal to the mass of the CO-core of the collapsing star with the 10\% gravitational mass defect. 
The nascent BHs are assigned an isotropic kick velocity $v_k=30\,\kms$. 
 (See for comparison with delayed collapse model \S~\ref{sec:disc}).

For the common envelope stages, we adopted the $\alpha$-formalism \citep{web84} with account 
for the binding energy of the envelope factor $\lambda$ \citep{dek90}, for which we use 
the values from \citet{2011ApJ...743...49L}. In the present study we assume the common envelope efficiency \ace=1.  

A critical issue in the formation of \lb1\ systems is the stability of mass transfer
from the companion to BH and the avoidance of merging of the components in the
common envelope which depends, mainly, on the mass and angular momentum loss
from the system. \citet{2017MNRAS.471.4256V}, using the fits to the evolutionary
tracks, have estimated that, if the mass ratio of the donor and BH is $\aplt$3.5
and the former has a radiative envelope, mass transfer through $L_1$ can be
stabilized by ``isotropic reemission''-- the loss of the excess of matter from
the system with the specific angular momentum of the accretor, e.g.,
\citet{sph97,2006LRR.....9....6P}. In addition, we have assumed that in
semi-detached systems with compact components $\delta$=10\% of the mass lost by
the donor leaves the system via the outer Lagrange point $L_2$ carrying specific
angular momentum of $L_2$. Binary separation change in this case is described,
e.g., by Eq.(6) in \citet{2018MNRAS.479.4844C}. This assumption almost does not
influence \mvis-\mbh\ relation, but affects the \porb-distribution of \lb1\
systems (see Figs.~\ref{fig:distr} and \ref{fig:distr_RC}). 

Instead of using formal criteria for the mass transfer stability and analytic
lifetime estimates of the stars of interest in the specific evolutionary stages,
as the second step of modelling we have applied stellar evolution code MESA
\citep[][and references therein]{2019ApJS..243...10P} to compute a sufficiently
dense grid of evolutionary tracks encompassing the \lb1 stage for the systems
with BHs taken from the BSE output. If a particular binary did not merge in the
common envelope, we used the computed lifetimes in the \lb1\ part of the track
to evaluate the Galactic number of such systems. Otherwise, we rejected the
binary from the statistics.

To calculate the current number of the Galactic \lb1 systems, we took into
account the star formation rate (SFR) in the Galactic thin disc after
\citet{2010A&A...521A..85Y}: $\mathrm{SFR}(t)=11e^{-(t-t_0)/\tau}+0.12(t-t_0)$,
with the time $t$ in Gyr, $t_0$=4 Gyr, $\tau=9$~Gyr, and the Galactic age of 
14~Gyr. This model gives the total mass of the Galactic thin disc $M_G=7.2\times
10^{10}$\,\msun, which we use for normalisation of calculations.

\section{Results}
\label{sec:results}
\begin{figure}
\label{fig:hrd}
     \includegraphics[width=0.5\textwidth]{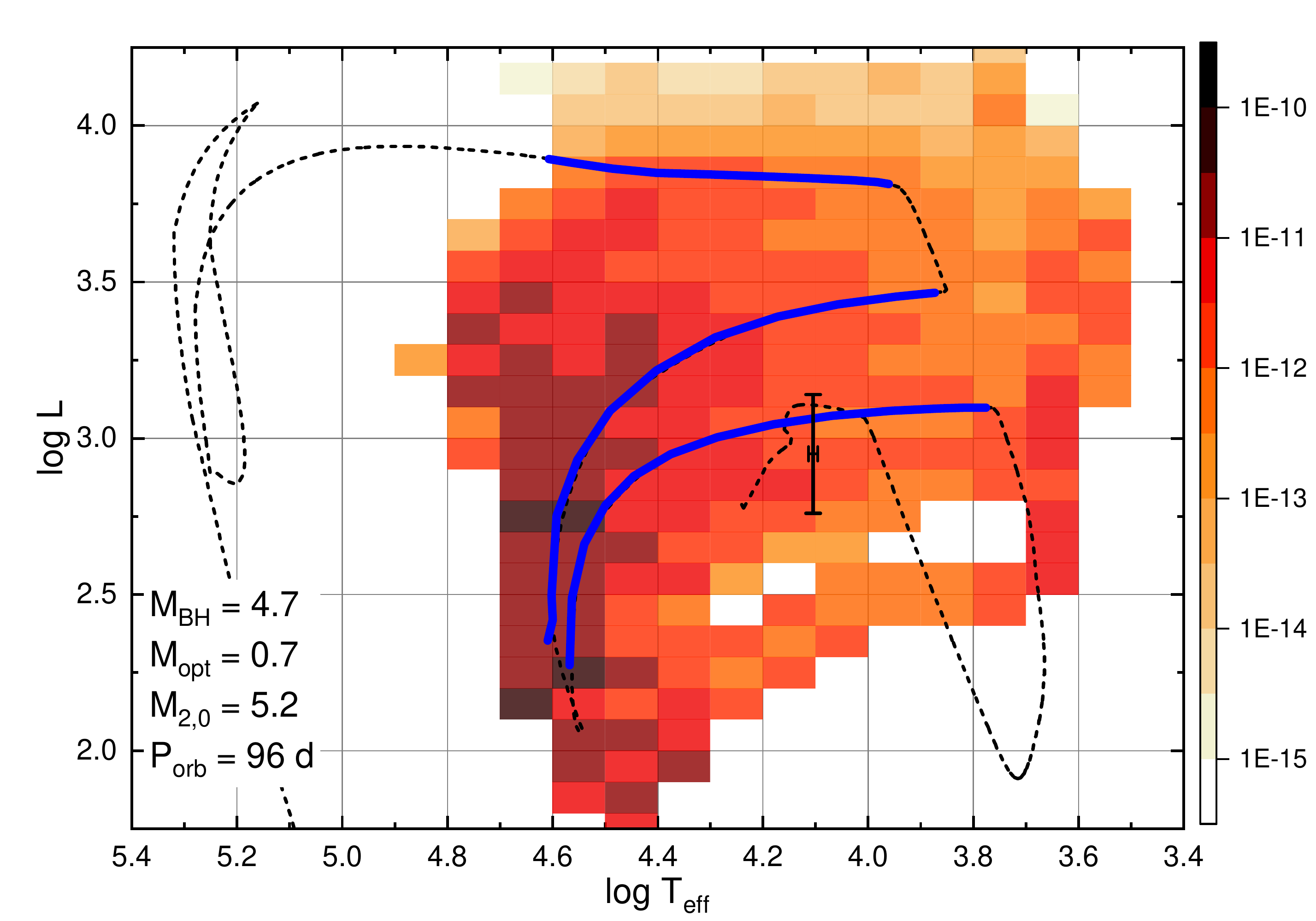}
   \caption{Model probability (per \ms) of finding \lb1\ systems  
   (colour scale) on HRD.  Overplotted is an evolutionary track of the visual component 
crossing HRD close to position of \lb1\ (error bars) suggested by \citet{2020A&A...633L...5I}.   
Thick solid lines mark H- or He-shell burning phases.} 
   
\end{figure}

\begin{figure}
   \centering   
\includegraphics[width=\columnwidth]{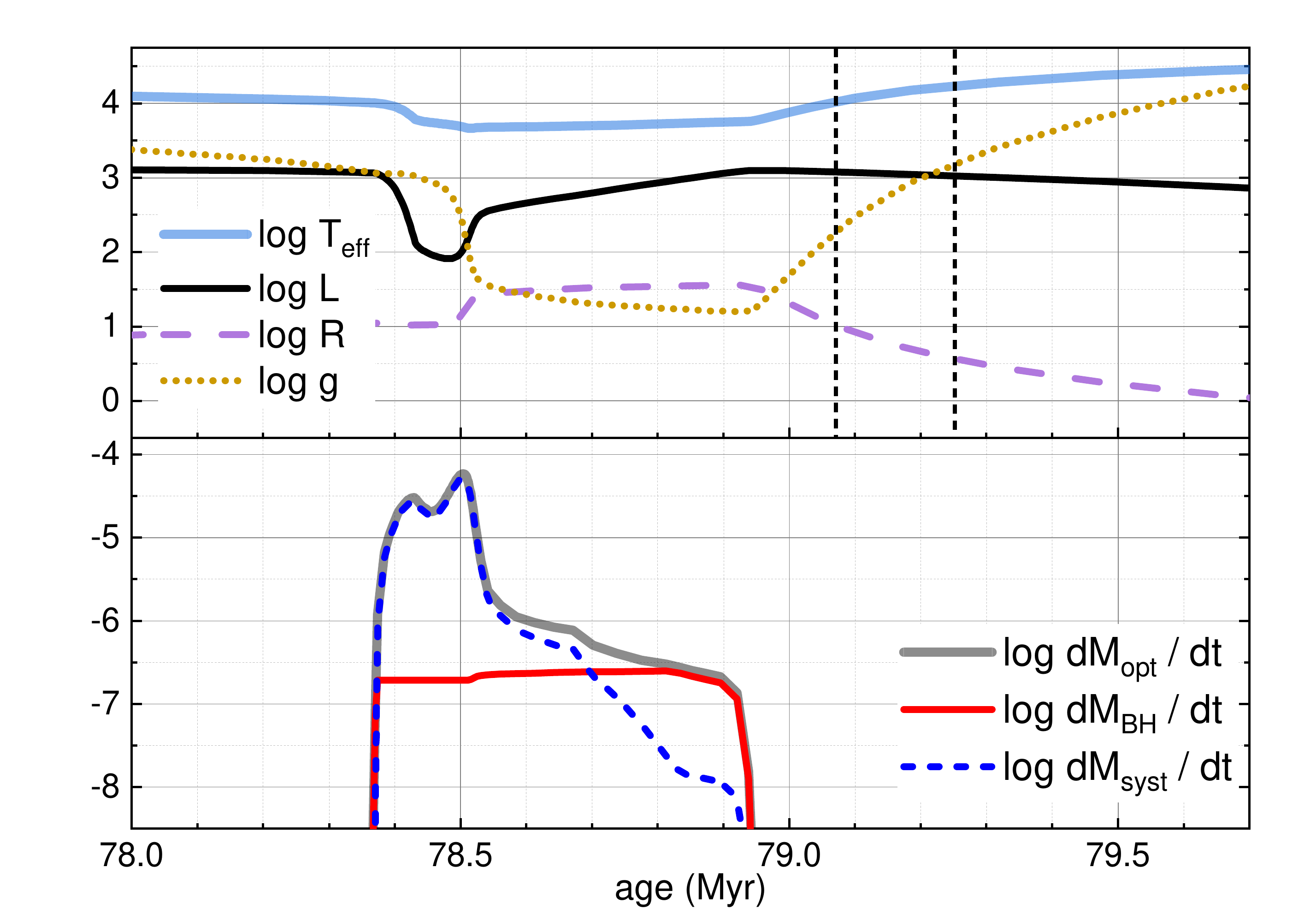}
   \caption{Upper panel:   parameters of the  optical component of the binary 
   shown in Fig.~\ref{fig:hrd} during RLOF and subsequent first transit of HRD. 
   Lower panel: donor's mass-loss rate, BH accretion rate, and mass-loss rate from the system during RLOF.  Vertical dotted lines mark the time span  when $\log(\teff)=4.0 - 4.2$. }
\label{fig:star}
\end{figure}

Figure~\ref{fig:hrd} presents the probability (in colour scale, per \ms) of
finding BHs with (0.5 - 1.7)\,\msun\ H- or He-shell burning companions, stripped
during case B of RLOF. We account the stars with luminosity of the H-burning
shell $L_H \geq L_{He}$, where $L_{He}$ is the luminosity of the He-burning
stellar core, and stars with CO-cores and He-burning shells. If they refill
Roche lobes, we do not count the duration of the RLOF phase in the statistics.
We assume that the \lb1\ stage ends when \teff\ of the star evolving in the HRD
diagram toward high \teff\ reaches that corresponding to the CO WD formation. In
fact, this location almost coincides with the ZAMS for He-stars. Note that the
probability of finding an \lb1\ system varies in the diagram within three orders
of magnitude and increases for high \teff\ because of slow-down of stellar
evolution with approach of He-core formation.

Overplotted in Fig. \ref{fig:hrd} is an evolutionary track for a star that
crosses HRD reasonably close to the position of \lb1\ and has at this time
\mvis, \mbh\ and  \porb\ similar to the ones reported by
\citet{2020A&A...633L...5I}\footnote{We did not attempt to tailor the track
which fits \lb1\ more precisely, since our main goal was to show that the
appearance and parameters of this type binaries can be naturally explained by
the state-of-the-art theory of stellar evolution.}. On ZAMS, the mass of BH
progenitor was 20.4\,\ms, \porb=202\,day. Before RLOF, \porb=5.8\,day, the orbit
was slightly eccentric (e=0.15). The binary crossed HRD after RLOF three times
in, resp., 2.2\,Myr, 7.0\,Myr, and 0.07\,Myr, i.e. the highest odds to observe
it as an \lb1\ system are when the optical component is in the He-shell burning
stage.

\begin{figure}
\includegraphics[width=\columnwidth]{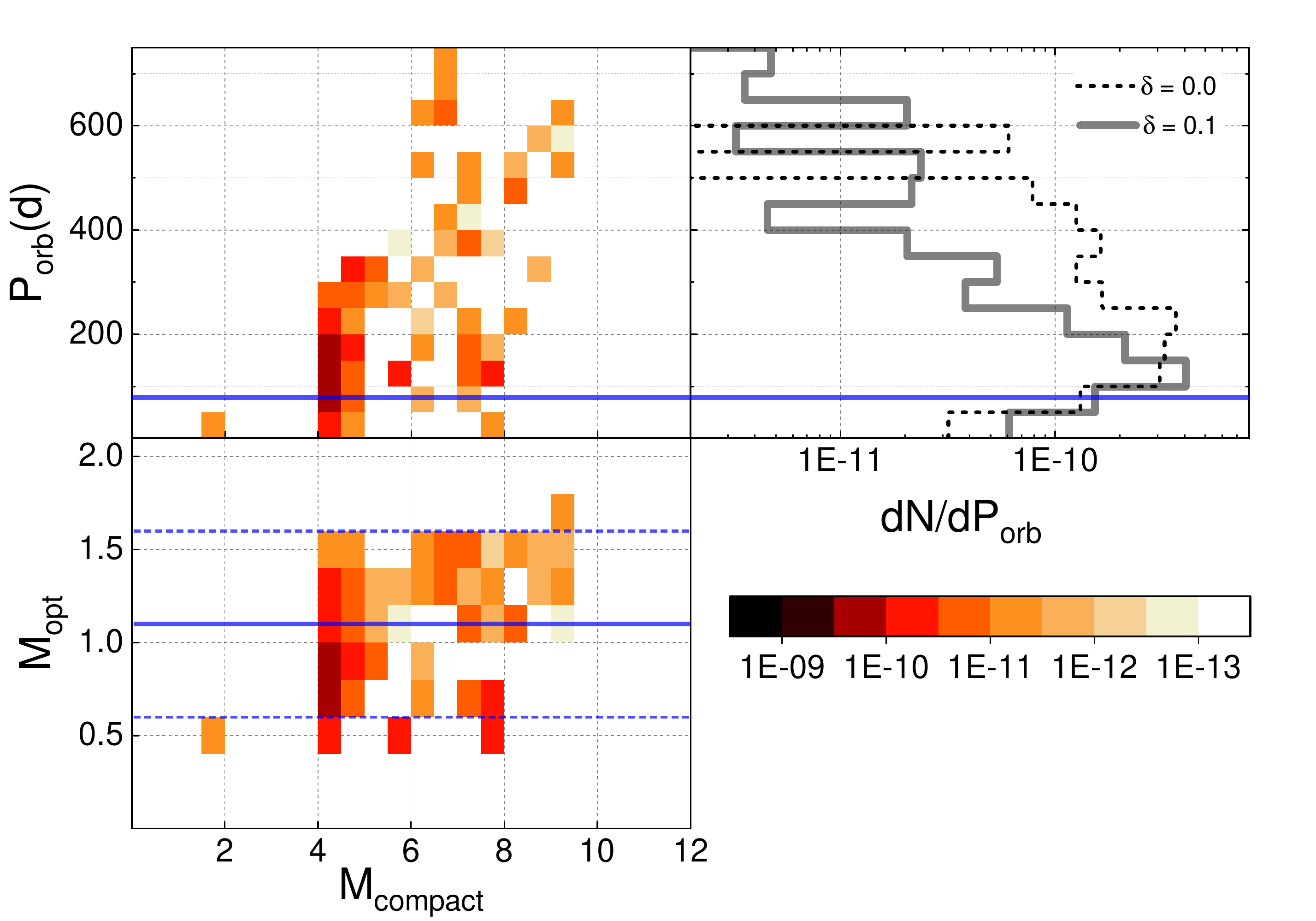}
   \caption{Distributions (per \ms) of combinations of component masses and orbital periods of \lb1\ type systems. 
   Dashed line shows distribution over periods if  no mass is lost via $L_2$. Blue lines show the range of estimates of $\mathrm{M_{opt}}$ and  \porb.}
 
   \label{fig:distr}
\end{figure}

The  upper panel of Fig. \ref{fig:star} shows the time dependence of the
parameters of the optical star in the binary shown in Fig.~\ref{fig:hrd} during
the RLOF stage and the first crossing of HRD, before formation of a He-star.
Note, the range of estimated \teff\ of \lb1\ is reached already in
$\simeq$5\,kyr after the end of RLOF. The presence of a gas disc around the
compact object in \lb1\ can be related to the brevity of this time. In the lower
panel of Fig. \ref{fig:star}, we plot the mass-loss rate by the donor, accretion
rate onto the BH, and mass-loss rate from the system during non-conservative
RLOF. Remarkably, the system avoids a common envelope despite
$\mathrm{(dM_{opt}/dt)}$ exceeds the Eddington limit by more than two orders of
magnitude. This means that the effect of the mass and momentum loss by the
isotropic re-emission leading to the widening of the system is much more
efficient than the opposite effect of the mass loss from $L_2$ (see also
Fig.~\ref{fig:distr}).

Figure~\ref{fig:distr} presents relations \mvis-\mbh\ and \mvis-\porb.
The plot reveals an evident trend of \mvis\ to the low values. The masses of their
progenitors are (3 -- 7)\ms. The masses of BH progenitors range from 18 to
40\,\ms. Note, the maximum of the \porb-distribution in the entire model
population is close to 100~day, not far from that of \lb1. However, we should
cautionary note that the stars that fit the range $\log(\teff)$=4.0-4.2 and have
the luminosity appropriate for \lb1\ itself comprise only $\simeq$1\% of the total
population and for them, under our assumptions, \porb$\simeq$50~day are more
typical. The plot clearly shows that the mass loss via $L_2$ plays a substantial
role for orbital periods of \lb1\ systems.

\begin{figure}
\includegraphics[width=\columnwidth]{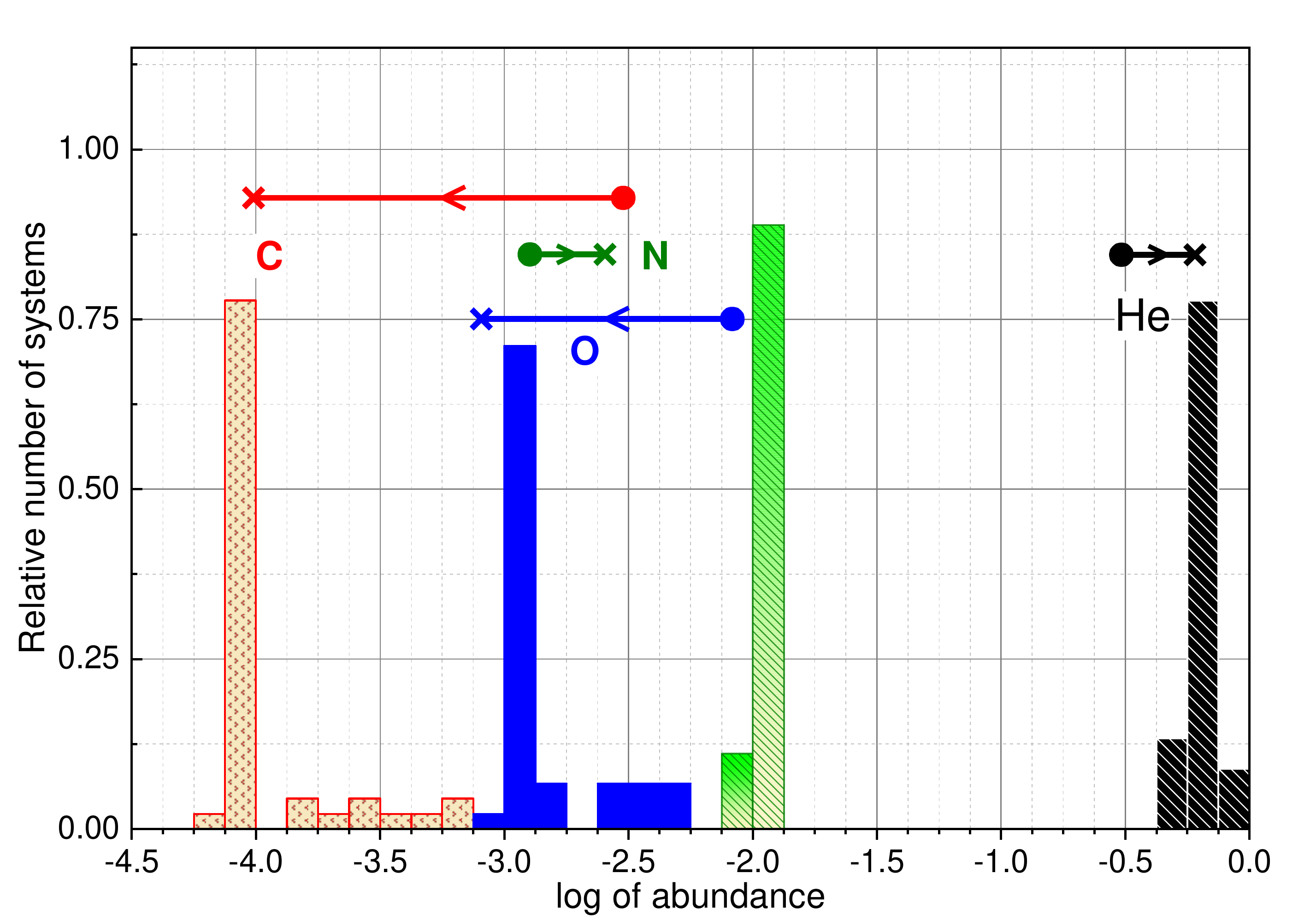}
   \caption{  
 Distribution of surface abundances of He, C, N, O 
of the model optical stars in LB-1 systems.  Horizontal lines connect abundances in homogeneous stars 
set in MESA (circles) and values  for LB-1 (crosses) derived by \citet{2020A&A...633L...5I}.
}
   \label{fig:chem}
\end{figure}

During the RLOF stage of the evolution of \lb1's precursors,
the interiors of their donors, where hydrogen burned, should be exposed.   The 
histogram of surface abundances of He, C, N, O in the models of donors after the end of  RLOF 
is shown in Fig.~\ref{fig:chem}. Later, 
the abundances hardly change, as the stellar winds from the stripped stars are expected to be very 
weak. These distributions 
should be compared to the initial abundances of the same
species assumed in MESA: 
$\log$(X(He))=-0.55, 
$\log$(X(C))=-2.46, 
$\log$(X(N))=-3.0, 
$\log$(X(O))=-2.03 
and the abundances derived  for \lb1\  \citep{2020A&A...633L...5I}:
$\log$(X(He))=-0.19,
$\log$(X(C))=-4.08, 
$\log$(X(N))=-2.53, 
$\log$(X(O))=-3.15 (we omit the errors).
{\citet{2020A&A...634L...7S} found depletion of C by a 
factor $\gtrsim 7$\ and enhancement of N by $\sim2.5$.
Examination of Fig.~\ref{fig:chem} and data of \citet{2020A&A...634L...7S} show
that the surface abundances of the model LB-1's optical components have the
trend expected for the layers of stars exposed after experiencing H-burning in
the CNO bi-cycle and loss of the envelopes in case B RLOF.} 

\begin{figure}
\includegraphics[width=\columnwidth]{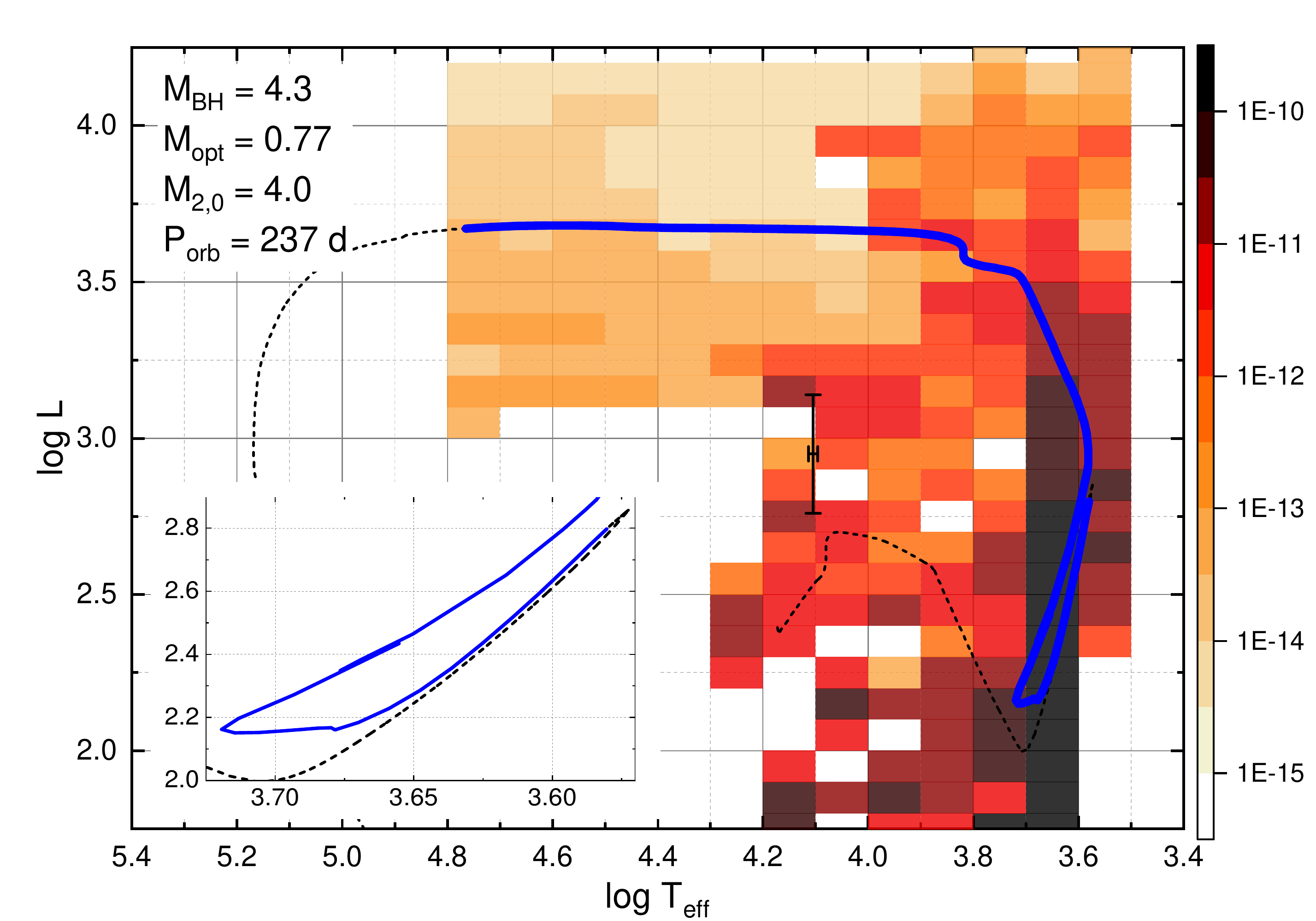}
   \caption{Model probability (per \ms) of finding 
   an RGB/AGB/post-AGB+BH system in HRD.  
  Blue lines correspond to the detached stages of post-RLOF donor evolution. }
   \label{fig:hrd_RC}
\end{figure}

\begin{figure}
\includegraphics[width=\columnwidth]{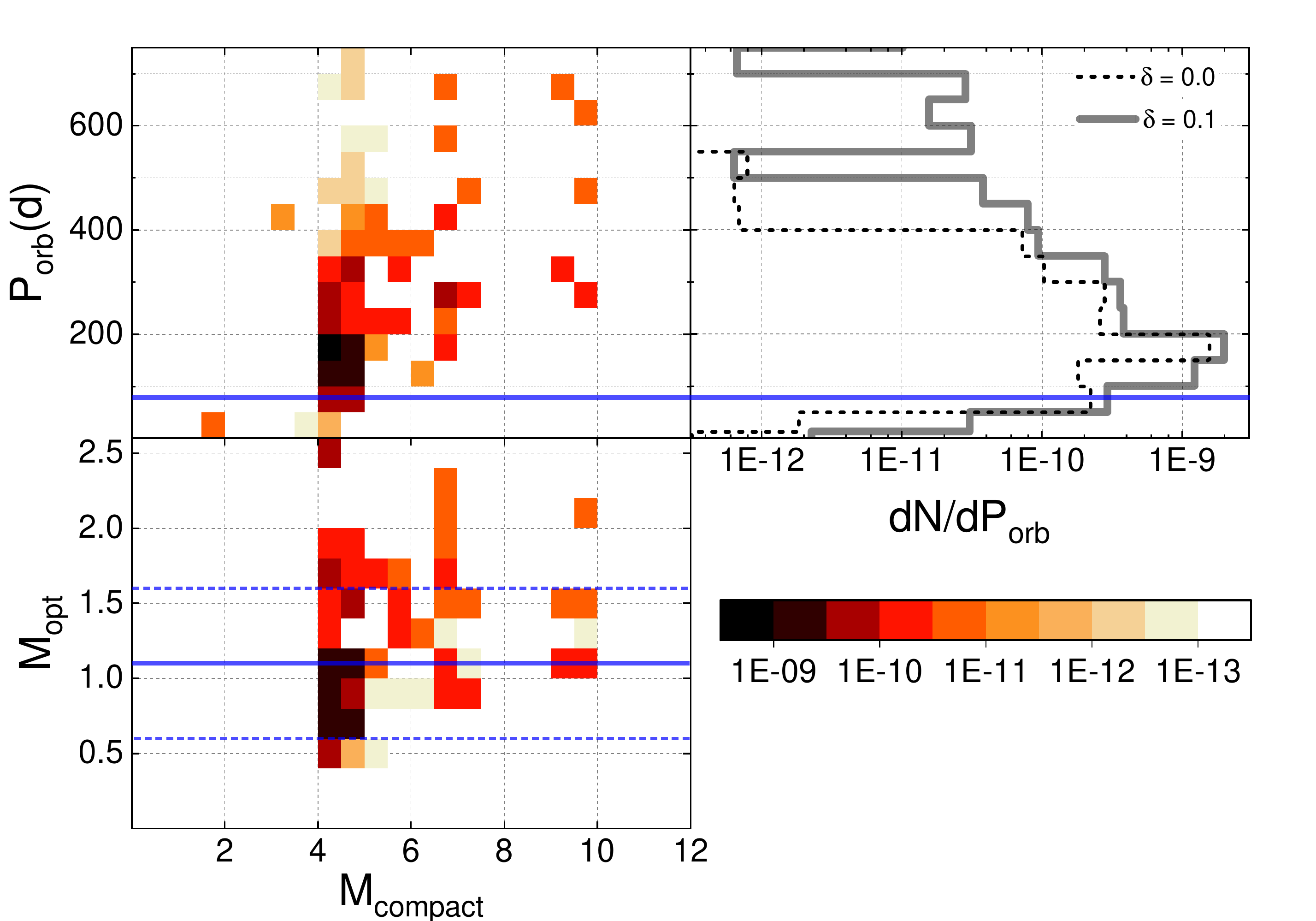}
   \caption{Same as in Fig.~\ref{fig:distr} for  RG/AGB + BH systems. }
   \label{fig:distr_RC}
\end{figure}
Convolution of the formation rate of \lb1's and related systems per \msun\ with SFR
in the Galactic disk enables an estimate of their current
Galactic number: $\simeq$200 core He-burning stars of (0.5-1.7)\,\ms\ with BH
companions, $\simeq$50 \lb1\ binaries burning H or He in the shells, and only a
few such stars with $\log(T_{\rm eff})$=4.0 - 4.2 and $\log(L/L_\odot)\geq2.8$.

We also found some binaries (typically, with pre-RLOF $M\aplt4$\,\ms, He-core
mass $\aplt0.5$\ms), which, upon completion of the RLOF, retain H-rich envelopes
of $\simeq$0.2\,\ms. They do not evolve to high \teff, but continue to move
along RG-branch in HRD, like single RGs with failed blue loops
(Figs.~\ref{fig:hrd_RC} and \ref{fig:distr_RC}). They may be called ``stripped
red giants''\footnote{An interesting problem, which is beyond the scope of our
study, is whether they may appear as symbiotic stars with BHs powered by wind
accretion from the RG/AGB companions.}. The end products of their evolution
should be ``hybrid'' CO/He WD. As post-AGB stars, they rapidly cross HRD, where,
in principle, they also may be discovered. We estimate their current Galactic
number as $\simeq 200$. There is also a small admixture of stars experiencing
case C RLOF (after core-He exhaustion) and later evolving up and leftward in the HRD. 
The
binary like 2MASS~J05215658+4359220, harbouring an RG and a massive unseen
component (\porb$\simeq$82\,day, $M_{RG}\simeq3.2\pm$1.0\,\ms,
$M_{comp}\simeq3.3^{+2.8}_{-0.7}$\,\ms, \citet{2019Sci...366..637T}), may be a
progenitor of this type stars. 

\section{Discussion and Conclusion} 
\label{sec:disc}
We have shown that the origin of binaries similar to \lb1\ and their parameters,
as derived by \citet{2020A&A...633L...5I}, can be explained by the standard
evolution theory of close binaries and current BHs formation scenarios. \lb1\ is
a star that has a very rapidly crossing HRD visual component that lost most of
H-rich mantle in the early case B RLOF and is in the H-shell burning stage. If the
mass of the visual component of the model star would exceed $\simeq$0.8\,\ms, He
in its core could be already exhausted and it could be in the He-shell burning
stage. Such stars may make several loops between ZAMS of He stars and the RGB
\citep[see, e.g.,][]{1985ApJS...58..661I} and have at that \teff\ and $L$
comparable to those of \lb1.
\lb1\ systems may have \porb\ in the range from several dozen to hundred day;
this range is strongly sensitive to the amount of mass and angular momentum lost
via $L_2$ during RLOF.

\citet{2020ApJ...890..113B}, using code MESA,  have shown that at $Z_\odot$ 
a 70\,\msun\ BH, in principle,  may form, if the 
stellar wind mass-loss rate is reduced by a factor $\sim$5 compared to the ``standard'' values \citep{1988A&AS...72..259D,2001A&A...369..574V}. 
However, any theoretical or observational justifications  for such a reduction of the  
mass loss are lacking.

\citet{2019arXiv191203599E}, as mentioned, computed a grid of models searching
for systems with parameters reported in \citet{2019Natur.575..618L}, but aiming
at the mass ratio \mbh/\mvis\ instead of the absolute value of \mbh. The ranges
\porb=(60 - 100)\,day and $\log(L_{\rm vis}/L_\odot)\geq3.0$ were attempted to
fit. Like in our study, \citet{2019arXiv191203599E} found that \lb1's may be
systems with stripped He-stars with H-burning shells and stellar mass (4 -
5)\,\msun\ BH components. However, the tracks they found, did not fit all
parameters of \lb1\ simultaneously.  

An important argument supporting the \lb1\ formation scenario via case B RLOF in
a binary, with an intermediate-mass secondary, as argued by
\citet{2020A&A...633L...5I}, is the consistency with the young age of \lb1,
inferred from its location in the Perseus arm. The short timescale of the
``transition'' of the putative progenitor of \lb1\ to its position in HRD after
RLOF (Fig.~\ref{fig:star}) allows one to consider the gas disc in the system as
a remnant of a structure formed during RLOF. This also explains the paucity of
\lb1\ systems.

We   have used some assumptions that need special mentioning. There is a
certain mismatch between the BSE and MESA calculations, since the former is
based on the the evolutionary tracks for single stars by
\citet{1998MNRAS.298..525P} and ``educated guesses'' for close binaries
evolution, while the latter is a full-fledged evolutionary code with the modern
input physics. Thus, the output of BSE modeling may differ from that of a BPS
code would it be based on MESA calculations. This  concerns mainly the masses of
stellar remnants, stellar lifetimes and the outcome of mass exchange. 

Above, we assigned to all BHs a uniform isotropic natal kick $v_k$=30\,. The problem
of BH formation and kicks is far from being solved. 
To test our assumptions on the BH kicks, in addition to the direct collapse model,  
we tested the BH formation model with fallback \citep{2012ApJ...749...91F}, as 
parameterized by
\citet[][Eq.~(8) and Appendix A]{2018MNRAS.474.2959G}. In this model, $v_k$ is randomly drawn from a Maxwellian distribution with one dimensional rms $\sigma=265\,\kms$, scaled by a function of the pre-SN masses of the star and its CO and Fe cores.
We found that the formation probability of BH+He-star binaries remained
qualitatively similar to the simpler case of the direct collapse. The total number of \lb1\ type systems  reduces by about 50\%.

LB-1's descend predominantly from the binaries with ZAMS $q \approx 0.2-0.4$. Other possible initial parameter distributions of massive  binaries or binarity 
rates,   e.g., 
$f(q) \propto q^{-0.1\pm0.6}$, $B=(0.69\pm0.09)$ \citep{2012Sci...337..444S}
or  $f(q) \propto q^{-1.7\pm0.3}$, $B=(1.0\pm0.2)$ \citep[][$B > 1$ means the presence of a close tertiary component]{2017ApJS..230...15M},
while being quite uncertain, can influence our estimates. Using the extreme exponents
in Sana et al. and Moe and Di Stefano, $f(q)$ changes the \lb1\ Galactic number by 
$\lesssim\pm 25\% $
Having in mind the uncertainty in the binarity $B$, we   conclude 
that our results depend on the statistical data within a factor of two, which is less than  other 
uncertainties plaguing BPS.
  
Assuming Z=0.02 for the entire lifetime of the Galactic disk, we estimate that there should 
be  $\sim2\times10^5$ wide non-merging BH+WD systems accumulated in the Galaxy. Stellar evolution is Z-dependent: in particular, stars with Z<$\mathrm Z_\odot$ are more compact and have more massive end products, thus affecting the occurrence
of case B mass exchange    
\citep[see, e.g.,][and references therein]{2020arXiv200400628K}. Hence, the formation of 
\lb1\ systems and their progeny is Z-dependent and should be traced along with the Galactic SFR 
and Z evolution. We do not consider such a task currently feasible due to the lack of systematic 
investigations of low-Z close binaries and the absence of statistical data on the distribution
over their masses, mass ratios and  orbital separations. However, it is possible  
that the lower formation rate of low-Z \lb1's may be compensated by a higher SFR, and the number 
of wide BH+WD could remain of the same order.

Despite the uncertainties of the evolutionary scenario we have used in our
study, we confirm that the origin of \lb1\ and the presence of a (4-10)\,\msun\
BH in this binary and in inevitably existing similar systems can be explained by
the standard stellar evolution which predicts the current presence of a few such
systems in the Galaxy.

\section*{Acknowledgements}
The authors thank the anonymous referee for insightful comments and U. Heber for useful discussion. 
This study was partially supported by RFBR grants 19-02-00790, 19-07-01198, 20-52-53009, 
and by the Leading Scientific School ``Physics of stars, relativistic objects and galaxies'' of Moscow University.


\bibliographystyle{mnras}
\bibliography{70msun_clean}
\bsp	
\label{lastpage}
\end{document}